\documentclass[sigconf]{acmart}
\usepackage{graphicx}
\usepackage{amsmath}
\usepackage{booktabs}
\usepackage{algorithm}
\usepackage{algorithmic}
\usepackage{amsfonts}
\usepackage{multirow}
\usepackage{makecell}
\usepackage{subfigure}
\usepackage{color}
\usepackage{bm}
\usepackage{epstopdf}
\usepackage{url}
\usepackage[cal=cm]{mathalfa}
\usepackage{balance}
\usepackage{threeparttable}
\usepackage{lipsum}
\usepackage{enumitem}
 \usepackage{pifont}
 \usepackage{tabularx}
 \usepackage{makecell}
 \usepackage{float}
 \usepackage{epigraph}

\usepackage{xcolor}
\newcommand{\myred}[1]{\textcolor{red}{#1}}
\newcommand{\mypurple}[1]{\textcolor{purple}{#1}}  

\AtBeginDocument{%
  \providecommand\BibTeX{{%
    \normalfont B\kern-0.5em{\scshape i\kern-0.25em b}\kern-0.8em\TeX}}}

\acmConference[CIKM '25]{The 34th ACM International Conference on Information and Knowledge Management}{November 10-14, 2025}{Seoul, Korea}
\acmYear{2025}
\acmMonth{11}

\renewcommand\footnotetextcopyrightpermission[1]{}

\begin{document}
\title{Pantheon: Personalized Multi-objective Ensemble Sort via\\Iterative Pareto Policy Optimization}

\author{Jiangxia Cao, Pengbo Xu, Yin Cheng, Kaiwei Guo, Jian Tang, Shijun Wang,\\Dewei Leng, Shuang Yang, Zhaojie Liu, Yanan Niu, Guorui Zhou, Kun Gai*}
\thanks{Kun Gai is the corresponding author.}
\affiliation{
  \institution{Kuaishou Technology, Beijing, China}
  \country{\{caojiangxia, xupengbo03\}@kuaishou.com, yin.sjtu@gmail.com, \{guokaiwei, tangjian03, wangshijun03, lengdewei, yangshuang08, zhaotianxing, niuyanan, zhouguorui\}@kuaishou.com, kun.gai@qq.com}
}

\renewcommand{\shorttitle}{Pantheon}

\begin{abstract}
RecSys engines have significantly advanced our daily-life, such as Kuaishou for short-video/live-streaming, Taobao for online-shopping, and so on.
To provide promising recommendation results, there exist three major stages in the industrial RecSys chain to support our service:
(1) The first \textbf{Retrieval model} aims at searching hundreds of item candidates.
(2) Next, the \textbf{Ranking model} estimates the multiple aspect probabilities \texttt{Pxtrs} for each retrieved item.
(3) At last, the \textbf{Ensemble Sort} stage merges those \texttt{Pxtrs} into one comparable score, and then selects the best dozen items with the highest scores to recommend them.
To our knowledge, the wide-accepted industry ensemble sort approach still relies on \textbf{manual formula-based adjustment}, i.e., assigning manual weights for \texttt{Pxtrs} to control its influence to generate the fusion score.
Under this framework, the RecSys severely relies on expert knowledge to determine satisfactory weight for each \texttt{Pxtr}, which blocks our system further advancements.

In this paper, we provide our milestone ensemble sort work and the first-hand practical experience, \textbf{Pantheon}, which transforms ensemble sorting from a "human-curated art" to a "machine-optimized science".
Compared with formulation-based ensemble sort, our Pantheon has the following advantages:
(1) \textbf{Personalized Joint Training}: our Pantheon is jointly trained with the real-time ranking model, which could capture ever-changing user personalized interests accurately.
(2) \textbf{Representation inheritance}: instead of the highly compressed \texttt{Pxtrs}, our Pantheon utilizes the fine-grained hidden-states as model input, which could benefit from the Ranking model to enhance our model complexity.
Meanwhile, to reach a balanced multi-objective ensemble sort, we further devise an \textbf{iterative Pareto policy optimization} (IPPO) strategy to consider the multiple objectives at the same time.
To our knowledge, this paper is the first work to replace the entire formulation-based ensemble sort in industry RecSys, which was fully deployed at Kuaishou live-streaming services, serving 400 Million users daily.
\end{abstract}

\ccsdesc[500]{Information systems~Recommender systems}

\keywords{Multi-Objective Optimization; Reinforcement Learning;}


\maketitle

\renewcommand{\shortauthors}{Jiangxia Cao, Pengbo Xu, Yin Cheng et al.}
\section{Introduction}
\begin{figure}[t!]
  \centering
  \includegraphics[width=9cm,height=5cm]{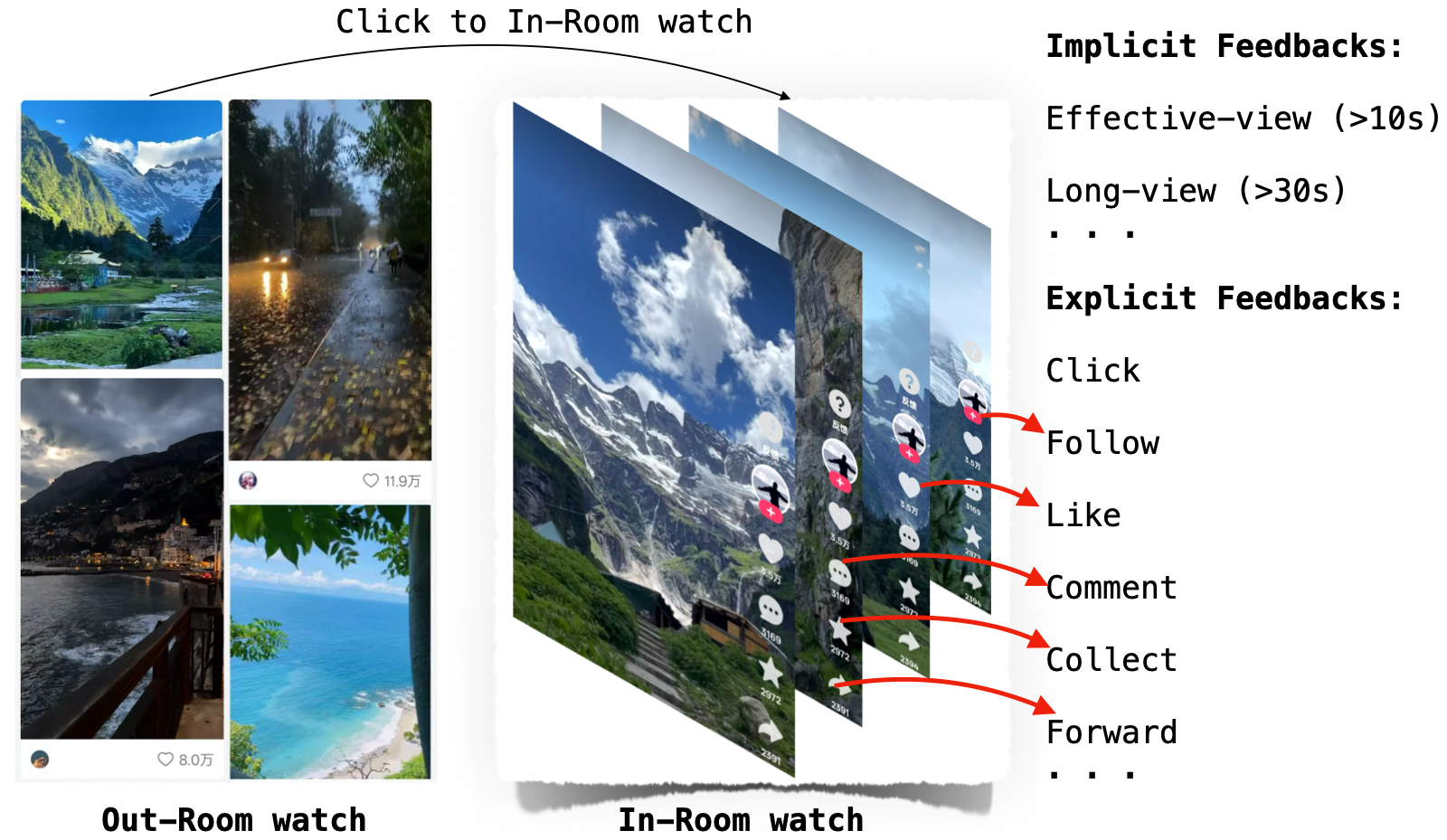}
  \caption{Multiple objectives at Kuaishou.}
  \label{kuaishou}
\end{figure}

\begin{figure*}[t!]
  \centering
  \includegraphics[width=18cm,height=6cm]{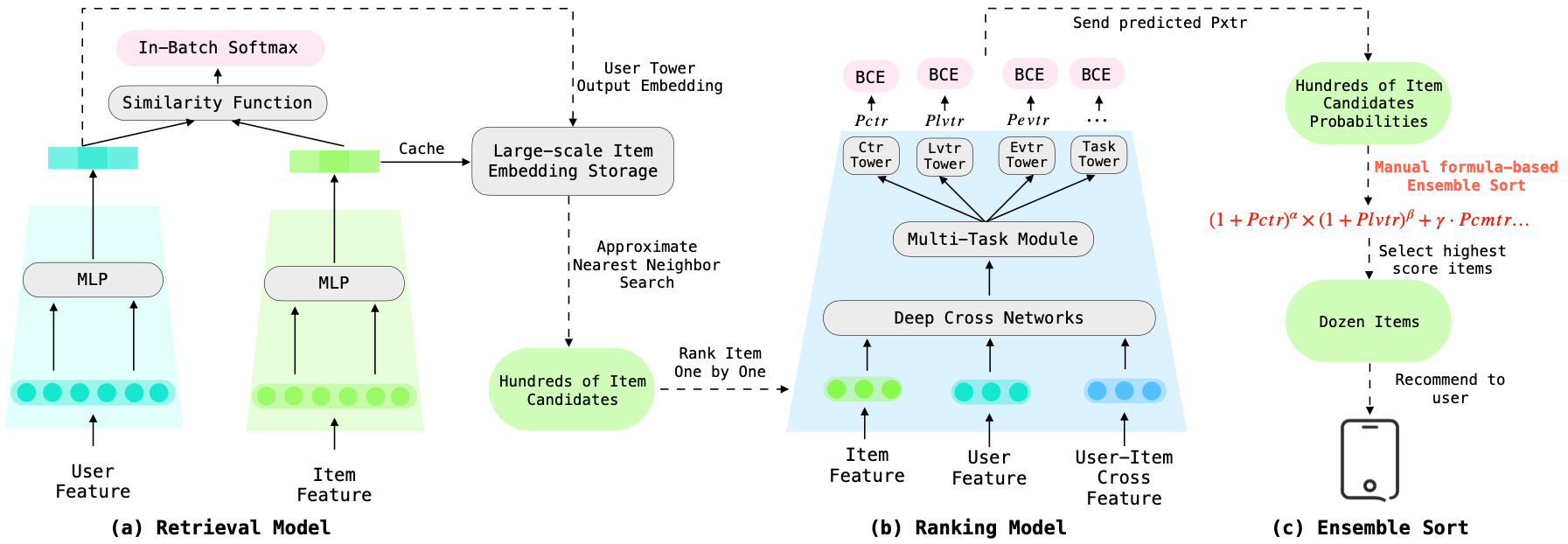}
  \caption{Illustration of RecSys chain: (a) the two-tower retrieval model searches hundreds of item candidates; (b) the ranking model to estimate the multiple objective probabilities; (c) the ensemble sort to fuse those prediction values as one score to select the top items.}
  \label{intro}
\end{figure*}

\setlength{\epigraphwidth}{.85\columnwidth}
\renewcommand{\epigraphflush}{center}
\renewcommand{\textflush}{flushepinormal}
\epigraph{\textit{``A system is enjoying maximum economic satisfaction when no one can be made better off without making someone else worse off."} }
{{\footnotesize{\textit{––Vilfredo Pareto}}}}

Kuaishou, as a leading short-video and live-streaming sharing platform, building a booming environment which attracts and benefits millions of users and creators worldwide.
As shown in Figure~\ref{kuaishou}, our Kuaishou offers a highly immersive experience through auto-play short-videos/live-streaming in full-screen mode by simply swiping Out-Room watching or clicking to In-Room watching, and leaving some interactions during surfing our platform~\cite{pepnet}, e.g., click, long-view, comment, etc.
To optimize user experience, our platform relies heavily on the recommendation system (RecSys) to ensure relevant content reaches the right users~\cite{liu2024crm}. 
However, with tens of millions of short-videos and live-streamings uploaded daily by creators, performing real-time scoring for every user-item pair across the entire space is computationally impossible~\cite{chen2024multi}.
To make a trade-off between effectiveness and efficiency, the industrial RecSys usually follows an elaborate cascading chain design paradigm with three major stages to respond user's recommendation request~\cite{mind, youtube}.
\begin{itemize}[leftmargin=*,align=left]
    \item User/Item information disentangled \textbf{Retrieval Model}~\cite{kuaiformer, dssm, dimerec}: as shown in Figure~\ref{intro}(a), the retrieval model always follow User/Item information disentanglement two-tower framework. 
    Therefore, we could easily find a small group of hundreds of item candidates from billion-scale item pool relies users' profile only.
    \item Multiple task learning guided \textbf{Ranking Model}~\cite{home, mmoe, ple}: as shown in Figure~\ref{intro}(b), the hybrid ranking model always mixes the user, item and user-item-cross information jointly to estimate multiple aspect interaction probabilities \texttt{Pxtrs} for retrieved items one by one, e.g., \texttt{ctr} of click rate, \texttt{lvtr} of long-view rate.
    \item Multiple objective optimizing oriented \textbf{Ensemble Sort}~\cite{sener2018multi}: as shown in Figure~\ref{intro}(c), the ensemble sort aims at fuse the predicted \texttt{Pxtrs} into a single comparable score. 
    Thus we could select the best dozen items with the highest scores to recommend them.
\end{itemize}

As the final stage in the recommendation pipeline, the Ensemble Sort~\cite{zhang2022multi} critically shapes both user experience and platform ecology to decide which content will be distributed to our users.
Therefore, the ensemble sort stage should be handled carefully handling, to find an optimal balance across multiple objective~\cite{haldar2023learning, kendall2018multi, tang2024multi}, such as total watch time, total clicks, and other metrics.
To the best of our knowledge, although there exists elaborate efforts~\cite{agarwal2019addressing,liu2023multitask,cao2025xmtf}, but most industrial RecSys still employ formula-based ensemble sorting mechanisms to measure final fused scores - typically using hybrid multiplicative or additive scoring formulas (in Figure~\ref{intro}(c)).
It is worth noting that there exist some weights (e.g., $\alpha$, $\beta$, $\gamma$) to control the importance of corresponding \texttt{Pxtr} in the fused score.
To find a group of suitable weights to maximize multiple objectives, the RecSys engineers could leverage offline parameter tuning tools~\cite{akiba2019optuna} (non-personalization, simple models with small data volume) with expert expertise to identify them.
However, in this setting, the highly flexible ("freestyle") nature of the formulation inherently caps its performance potential. It often demands excessively complex rule engineering to reach a Pareto-optimal trade-off among user experience metrics — \textit{where no metric can be improved without degrading another}.

At Kuaishou, our former ensemble sort mechanism is also equipped by a complex formulation, which has been iterated for several years.
While delivering significant initial benefits, the limitations of the pre-defined style formula have become increasingly apparent over time. 
Meanwhile, iteration requires a heavy reliance on manual expertise and extensive A/B testing, resulting in deterioration in efficiency.
Such "hand-crafted" ensemble sort has become a bottleneck hindering further advancements in our system, preventing the full potential of massive data~\cite{lv2024marm} and complex models scaling~\cite{gpt4}.
To this end, we started to explore the neural-network based ensemble sort - towards to find a new way to unleash it from a "human-curated art" to a "machine-optimized science".
Considering that all \texttt{Pxtrs} metrics are generated from the ranking model (in Figure~\ref{intro}), could we develop a `plugin' module within the ranking model to fuse them?

Motivates by this, we propose our first-hand practical milestone work, \textbf{Pantheon},  which \textbf{successfully replace the traditional formulation-style ensemble sort mechanism in our system}.
In our model architecture designing, our Pantheon has the following advantages compared with formulation-based ensemble sort:
\begin{itemize}[leftmargin=*,align=left]
\item Highly \textbf{personalized} joint training: In the fusion process, our Pantheon jointly trains with the real-time ranking model~\cite{cao2024moment} and assigns un-shared learnable user and item features as additional information to generate the ensemble score, achieving granular personalization through adaptive ensemble.
\item Fine-grained \textbf{representation inheritance}: Actually, propagating these few numerical scores \texttt{Pxtr} leads to severe information decay. Here we reuse the high-dimensional task-specific representations from the task tower as our model input~\cite{ple}, which could benefit our ensemble sort from the computationally complex Ranking model, to enhance our model effectiveness.
\end{itemize}
In the label side, we follow the standard simple additive weighting mechanism to combine the multiple objectives of user experience, e.g., $\sum_{\texttt{Pxtr}}^{\{\texttt{Pctr}, \texttt{Plvtr}, \dots\}}w^{\texttt{Pxtr}}\mathcal{L}^{\texttt{Pxtr}}$.
Supervised by the additive function, our model is encouraged to converge to a locally optimal point theoretically, while the quality of the `locally convergence' critically depends on the weighting coefficients $w^{\texttt{Pxtr}}$ of these objectives.
To guarantee convergence to Pareto-optimal states~\cite{desideri2012multiple, kurin2022defense, ma2020efficient}, we design an Iterative Pareto Policy Optimization (IPPO) mechanism from a reinforcement learning (RL) perspective~\cite{liu2023multi}, to automatically search a group of weights to reach a Pareto frontier.

In summary, our contributions are as follows:
\begin{itemize}[leftmargin=*,align=left]
\item We propose our neural-network based ensemble sort approach, to our knowledge, Pantheon is the first work to replace the entire formulation-based ensemble sort in industry RecSys. 
\item We offer insights into both model architecture and training policy, addressing practical challenges encountered in industrial RecSys, which will shed light on other researchers to explore a more robust ensemble sort.
\item We conduct extensive offline and online experiments to verify our Pantheon effectiveness, which contributes more than $1\%$ of Clicked User online gains to Kuaishou live-streaming services.
\end{itemize}

\section{Preliminary}
In this section, we briefly review: (1) The multi-task learning based Ranking model in RecSys to produce $\texttt{Pxtrs}$; (2) the traditional \texttt{Pxtr} based ensemble sort for multi-objective optimization~\cite{fliege2000steepest, lin2019pareto} in RecSys.

\subsection{Multi-Task Learning in RecSys}
In practice, the multi-task learning always conducted at ranking model, which includes the following four major components:
\begin{itemize}[leftmargin=*,align=left]
\item \textbf{Training Label}: During watching live-streamings, users always leave a large amount of interaction logs, such as click, long-view, effective-view, and others.
Thus, each (user, item) view will generate multiple ground-truth signals, e.g., click $y^{\texttt{ctr}}\in \{0, 1\}$, long-view $y^{\texttt{lvtr}}\in \{0, 1\}$, effective-view $y^{\texttt{evtr}}\in \{0, 1\}$, $\dots$.
\item \textbf{Input Feature}: Generally speaking, the hybrid ranking model integrates user, item, and user-item cross-features, which can be roughly divided into four categories:
(1) ID features: user ID, item ID, category ID, etc, 
(2) Statistical features: user watched live-streaming in the last month, live-streamingmetrics~\cite{lu2025liveforesighter} total watched times, to describe user activity levels or item popularity.
(3) Sequence features: user recent or searched live-streaming sequences, such as DIN~\cite{din}, SIM~\cite{sim}. 
(4) Multi-modal features: LLM-generated item content embedding and Semantic ID, such as LARM~\cite{liu2025llm}, QARM~\cite{luo2024qarm}.
\item \textbf{Model Architecture}: For brevity, let denote the above \textbf{input features as $\mathbf{v}$}, the multi-task learning can be formed as:
\begin{equation}
\small
\begin{split}
 \mathbf{e}^{\texttt{ctr}}, \mathbf{e}^{\texttt{lvtr}}, \dots &= \texttt{Mixture-of-Expert}(\mathbf{v}),\\
 \mathbf{t}^{\texttt{ctr}} = \texttt{Tower}^{\texttt{ctr}}(&\mathbf{e}^{\texttt{ctr}}), \ \ \mathbf{t}^{\texttt{lvtr}} = \texttt{Tower}^{\texttt{lvtr}}(\mathbf{e}^{\texttt{lvtr}}), \ \ \dots,\\
 \hat{y}^{\texttt{ctr}} = \texttt{Pred}^{\texttt{ctr}}(&\mathbf{t}^{\texttt{ctr}}), \ \ \hat{y}^{\texttt{lvtr}} = \texttt{Pred}^{\texttt{lvtr}}(\mathbf{t}^{\texttt{lvtr}}), \ \ \dots,\\
\end{split}
\label{moe}
\end{equation}
where \myred{the $\mathbf{t}^{\texttt{ctr}}, \mathbf{t}^{\texttt{lvtr}}\in \mathbb{R}^d$ are the tasks' hidden-states, and the $\hat{y}^{\texttt{ctr}}, \hat{y}^{\texttt{lvtr}}\in (0,1)$ are the predicted scores}. The $\texttt{Pred}(\cdot)$ are single-layer \texttt{MLP} with \texttt{Sigmoid} activated function, the $\texttt{Tower}(\cdot)$ are stacked \texttt{MLP} with \texttt{ReLU} activated function, and the multi-task module $\texttt{Mixture-of-Expert}(\cdot)$ is a gate-expert paradigm networks, such as the MMoE~\cite{mmoe}, PLE~\cite{ple} or HoME~\cite{home}.
The $\hat{y}^{\texttt{ctr}}\in(0,1)$ and $\hat{y}^{\texttt{lvtr}}\in(0,1)$ are the predicted probabilities, i.e., \texttt{Pxtr}.
\item \textbf{Loss Function}: According to the predicted \texttt{Pxtr} $\{\hat{y}^{\texttt{ctr}}, \hat{y}^{\texttt{lvtr}}, \dots \}$ and the ground-truth label $\{y^{\texttt{ctr}}, y^{\texttt{lvtr}}, \dots \}$, the ranking model conduct multiple binary cross-entropy classification loss to optimize different \texttt{Pxtr}.
\begin{equation}
\small
\begin{split}
\mathcal{L}^{\texttt{ctr}}_{\texttt{ranking}} = - &\big(y^{\texttt{ctr}}\log{(\hat{y}^{\texttt{ctr}})} - (1-y^{\texttt{ctr}})\log{(1-\hat{y}^{\texttt{ctr}}})\big), \\
\mathcal{L}^{\texttt{lvtr}}_{\texttt{ranking}} = - &\big(y^{\texttt{lvtr}}\log{(\hat{y}^{\texttt{lvtr}})} - (1-y^{\texttt{lvtr}})\log{(1-\hat{y}^{\texttt{lvtr}}})\big), \\
 \mathcal{L}_{\texttt{ranking}} = & \mathcal{L}^{\texttt{ctr}}_{\texttt{ranking}} + \mathcal{L}^{\texttt{lvtr}}_{\texttt{ranking}} + \dots
\end{split}
\label{rank_bce}
\end{equation}
\end{itemize}
where the $\mathcal{L}_{\texttt{ranking}}$ is the final training loss.
Since the different objectives assigned different prediction tower, \textbf{the ranking model could achieve the maximizing accuracy for each objective}.

\subsection{Multi-Objective Optimization in RecSys}
After the ranking model, the multi-objective optimization~\cite{lin2025alignpxtr} aims at selecting the best items that have the highest score for user experience.
Here we review an \textbf{offline}  parameter searching workflow for the formula-based ensemble sort:
\begin{itemize}[leftmargin=*,align=left]
\item \textbf{Offline Data Collection}: In an offline setting, we first collect millions of data samples, while each samples records the ranking model predictions results \texttt{Pxtrs} and the ground-truth labels i.e., $y^{\texttt{ctr}}\in \{0, 1\}$, long-view $y^{\texttt{lvtr}}\in \{0, 1\}$, and so on.
\item \textbf{Pre-Defined Formulation Style}: Based on these samples, we then copy the pre-defined online base ensemble sort fusion formula style. Here we provide a toy example:
\begin{equation}
\begin{split}
\texttt{Score} = (1 + \hat{y}^{\texttt{ctr}})^\alpha\times(1 + \hat{y}^{\texttt{lvtr}})^\beta + \gamma\cdot\hat{y}^{\texttt{evtr}} + \dots
\end{split}
\label{formulation}
\end{equation}
where the \texttt{Score} is the fusion score of each sample, and $\alpha, \beta, \gamma$ are formula parameters.
\item \textbf{Hand-Crafted Evaluation Metric}: During parameter searching, we need to hand-craft an \textbf{intermediate evaluation target to reflect our goal}, such as ensuring the final aggregated \texttt{Score} accurately represents both click-through and long-play precision.
\begin{equation}
\begin{split}
\texttt{EvalMetric} = & 2\times\texttt{GAUC}(\texttt{Score}, y^{\texttt{ctr}}) \\
& + 5 \times \texttt{GAUC}(\texttt{Score}, y^{\texttt{lvtr}}) + \dots
\end{split}
\label{evaluation}
\end{equation}
In contrast to ranking models, here we employ the fused \texttt{Score} for all objectives to calculate the GAUC (see the details in Sec.\ref {gauc_exp}). 
The fixed hyper-parameters \textbf{2 and 5 are according to the expert knowledge, which should be carefully selected}.
Actually, these weights reflect the relative business importance with the expert knowledge.
\item \textbf{Parameter Tuning}: Based on the pre-defined formulation style and hand-crafted evaluation metric, our goal is to search for a group of appropriate weights which maximize the \texttt{EvalMetric}:
\begin{equation}
\begin{split}
\alpha^*, \beta^*, \gamma^* = \mathrm{argmax}_{\alpha, \beta, \gamma} \texttt{EvalMetric}
\end{split}
\label{searching}
\end{equation}
\end{itemize}
where the $\alpha^*, \beta^*, \gamma^* \in \mathbb{R}$ are the optimal parameters under the \texttt{EvalMetric} setting.
In this way, we typically employ Bayesian-based parameter search tools to accelerate this process, such as TPE-based Optuna and Paradance\footnote{\url{https://github.com/optuna/optuna} and \url{https://github.com/yinsn/ParaDance}. Specifically, the ParaDance is an open-source tool developed by this paper author Yin Cheng.}.
However, the searched parameters may \textbf{only achieve optimal performance at the single-dimensional \texttt{EvalMetric}} in the offline datasets, not necessarily reflect the real-world online A/B test performance from diverse aspects.
Therefore, we typically combine the parameter searching with expert knowledge to ensure online effectiveness with further hand-crafted efforts.

\begin{figure*}[t!]
  \centering
  \includegraphics[width=18cm,height=6.5cm]{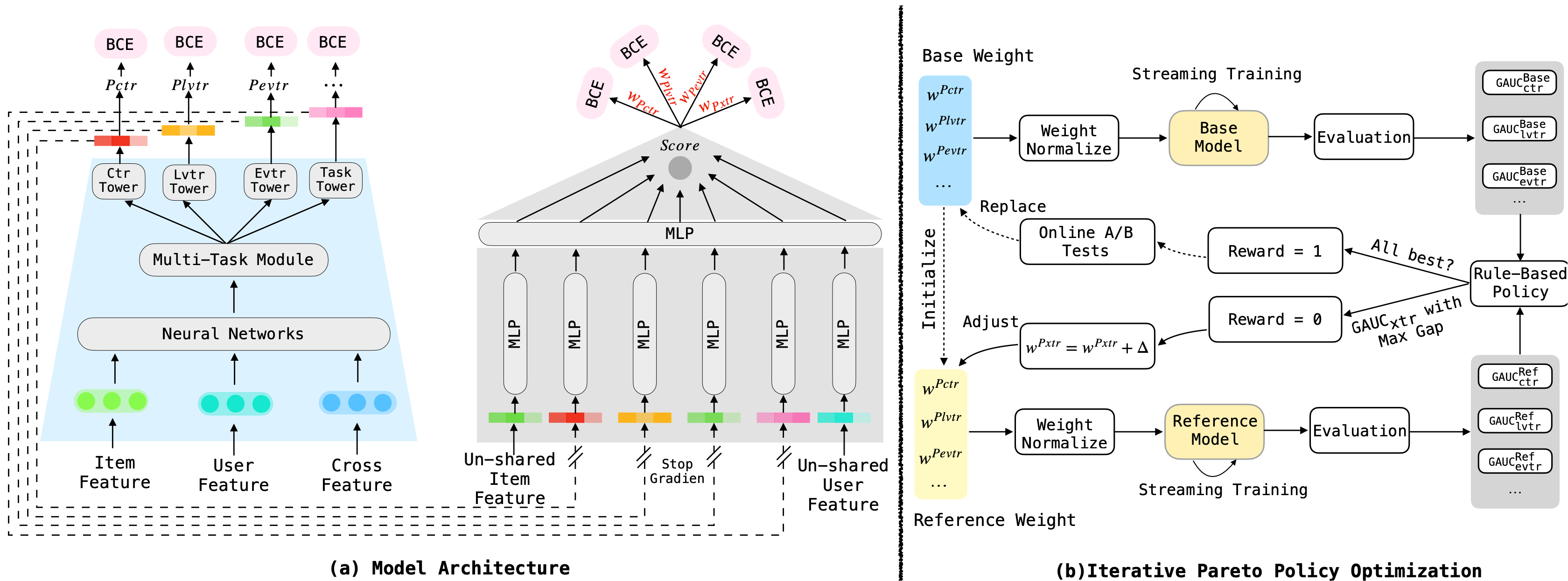}
  \caption{Pantheon model architecture and its iterative Pareto policy optimization strategy.}
  \label{Pantheon}
\end{figure*}

\section{Pantheon Workflow}
In this section, we introduce the details of our model, Pantheon.
We first express how our Pantheon performs multi-objective optimization through joint training with the ranking model.
Afterwards, we describes the IPPO strategy to show how we find a group of loss weights meets the Pareto-optimal constraints.

\subsection{Fusion Score Generation}
As shown in Figure~\ref{Pantheon}(a), our Pantheon can be seen as an additional ‘plugin’ component from ranking model.
Here we offer our model architecture and training loss designing insights.

\subsubsection{Model Architecture}
Benefiting from joint training with the ranking model, our training framework has the advantages of large-scale parameters and massive training data naturally.
Hence, in model designing, we focus on that: (a) reuse the extracted high-dimension characterization from sophisticated ranking model, we concatenate the multiple tasks' representations as our Pantheon inputs, which alleviates the information decay of highly-compressed \texttt{Pxtr}  (b) to avoid two-staged formulation-based ensemble sort paradigms, build an end-to-end ensemble model.
For the first point, while employing the high-dimensional task-specific representations $\mathbf{t}$ (in Eq.(\ref{moe})) to support our Pantheon, we apply \myred{the `stop-gradient' operator to disentangle} our ensemble sort component with the ranking model, which enforces gradient isolation between the ranking model and Pantheon.
For the second point, we \myred{introduce an additional set of learnable user/item features} to serve as personalized inputs for our Pantheon.
According to them, our Pantheon inputs can be formed as:
\begin{equation}
\small
\begin{split}
& \mathbf{P} = \{\mathbf{ItemFea}, \ \texttt{Stop-Gradient}(\mathbf{t}^{\texttt{ctr}}),\ 
\texttt{Stop-Gradient}(\mathbf{t}^{\texttt{lvtr}}),\\
&\texttt{Stop-Gradient}(\mathbf{t}^{\texttt{cmtr}}),\dots,
\texttt{Stop-Gradient}(\mathbf{t}^{\texttt{xtr}}), \mathbf{UserFea}\}.
\end{split}
\label{input}
\end{equation}
where the $\mathbf{ItemFea}$/$\mathbf{UserFea}\in \mathbb{R}^d$ are the un-shared additional item/user features, $\mathbf{t}^{\texttt{xtr}}\in \mathbb{R}^d$ are the tower output representation in Eq.(~\ref{moe}), and the $\mathbf{P}$ denotes the high-dimensional final input of our Pantheon.
In our experiments, we find that the un-shared additional item/user features also significantly accelerated the convergence of Pantheon.
According to the fine-grained information, we utilize a simple network to generate the fusion score as:
\begin{equation}
\begin{split}
\texttt{Score} = \texttt{Ensemble\_Encoder}(\mathbf{P}).
\end{split}
\label{ensembleencoder}
\end{equation}
where the $\texttt{Ensemble\_Encoder}(\cdot)$ is a \texttt{MLP}-based networks with \texttt{Sigmoid} activated function (in Figure~\ref{Pantheon}(a)), and the $\texttt{Score}\in (0, 1)$ is the final fused float result.

\subsubsection{Training loss function}
Compared with original ranking model that utilizes different $\texttt{Pxtr}$ to predict different tasks in Eq.(~\ref{rank_bce}), in multi-objective optimization paradigm, we need to apply a single Score to compare different items while maximizes the multiple objectives metrics.
Therefore, in our Pantheon, we employ a standard simple additive weighting mechanism to combine multiple objectives information:
\begin{equation}
\small
\begin{split}
\mathcal{L}^{\texttt{ctr}}_{\texttt{Pantheon}} = - &\big(y^{\texttt{ctr}}\log{(\texttt{Score})} - (1-y^{\texttt{ctr}})\log{(1-\texttt{Score}})\big), \\
\mathcal{L}^{\texttt{lvtr}}_{\texttt{Pantheon}} = - &\big(y^{\texttt{lvtr}}\log{(\texttt{Score})} - (1-y^{\texttt{lvtr}})\log{(1-\texttt{Score}})\big), \\
 \mathcal{L}_{\texttt{Pantheon}} = & w^{\texttt{ctr}}\mathcal{L}^{\texttt{ctr}}_{\texttt{Pantheon}} + w^{\texttt{lvtr}}\mathcal{L}^{\texttt{lvtr}}_{\texttt{Pantheon}} + \dots
\end{split}
\label{loss_func}
\end{equation}
where the $\mathcal{L}_{\texttt{Pantheon}}$ is the final training loss objective, and the $w^{\texttt{ctr}}$, $ w^{\texttt{lvtr}}$ are the \textbf{positive weights} to balance different objectives importance. 
Since different objectives show different positive sample densities, the influence of different tasks varies even same weight. 
To avoid our model degenerate into a pure click model, we employ such weights to balance the influence of tasks, which can lead to learn the global objectives' interest of each user.
Particularly, in our experiments, we found that the weight of the loss is crucial to the final convergence result of the model. In the next section, we will elaborate on our IPPO strategy to show the process of how to iterate these weights.

\subsection{Iterative Pareto Policy Optimization}
In this section, we give our training policy from reinforcement learning perspective, to automatically search a group of weights to reach a Pareto frontier.

\subsubsection{Basic Concepts}
Before going on, we first introduce some basic concepts for better understanding:
\begin{proposition}
\textbf{Pareto Optimality under \myred{Positive Weights}}. 
Given scalarized loss $\mathcal{L} = \sum_{i=1}^n w^i \mathcal{L}^i$ with strictly positive weights $w^i > 0\ (\forall i)$, let $\theta^*$ be a local minimizer of $\mathcal{L}$. Then $\theta^*$ is locally Pareto optimal for the multi-objective problem $\min_{\theta} (\mathcal{L}^1, \dots, \mathcal{L}^n)$. Formally:
\begin{equation}
\theta^* \in \mathop{\mathrm{arg\,min}}_\theta \mathcal{L} \implies \nexists \theta' \text{ s.t. } 
\begin{cases} 
\mathcal{L}^i(\theta') \leq \mathcal{L}^i(\theta^*),\ \forall i \\
\mathcal{L}^k(\theta') < \mathcal{L}^k(\theta^*),\ \exists k 
\end{cases}
\label{po}
\end{equation}
\end{proposition}

\textit{Proof}: Assume the contrary that $\exists \theta'$ Pareto dominating $\theta^*$. Then:
\begin{equation}
\begin{split}
\sum_{i=1}^n w^i \mathcal{L}^i(\theta') & < \sum_{i=1}^n w^i \mathcal{L}^i(\theta^*) \quad (\because w^k > 0 \text{ and } \mathcal{L}^k(\theta') < \mathcal{L}^k(\theta^*)) \\
& \Rightarrow \mathcal{L}(\theta') < \mathcal{L}(\theta^*)
\end{split}
\end{equation}
contradicting the local optimality of $\theta^*$ for $\mathcal{L}$. \hfill $\square$
\begin{proposition}
\textbf{Invariance Property under Homogeneous Scaling}. For any positive constant $k>0$, rescaling the total loss $\mathcal{L} = \sum_{\texttt{xtr}}^{\{\texttt{ctr}, \texttt{lvtr}, \dots\}}w^{\texttt{xtr}}\mathcal{L}^{\texttt{xtr}} \xrightarrow{} k\mathcal{L} = \sum_{\texttt{xtr}}^{\{\texttt{ctr}, \texttt{lvtr}, \dots\}}kw^{\texttt{xtr}}\mathcal{L}^{\texttt{xtr}}$ preserves the optimal model parameters and the resulting score distribution. Formally, $\theta^*$ be the same optimal model parameters:
\begin{equation}
\begin{split}
\theta^* = \mathrm{argmin}_\theta\ \mathcal{L} = \mathrm{argmin}_\theta\ k\times \mathcal{L}
\end{split}
\label{optimial}
\end{equation}
\end{proposition}
\textit{Proof}: We first derive the gradient direction as:
\begin{equation}
\begin{split}
\nabla_\theta(k\mathcal{L}) = k\nabla_\theta\mathcal{L} \propto \nabla_\theta(\mathcal{L}) = \nabla_\theta\mathcal{L} 
\end{split}
\label{gradient}
\end{equation}
Thus, gradient descent paths for $\mathcal{L}$ and $k\mathcal{L}$ are identical when learning rates are scaled by $\frac{1}{k}$.
Next, at the optimal convergence status, both losses share the same critical point condition:
\begin{equation}
\begin{split}
k\nabla_\theta\mathcal{L} = 0 \iff \nabla_\theta\mathcal{L} = 0
\end{split}
\label{optimal}
\end{equation}
Thus \myred{weight normalization does not change the convergence point}.
\begin{definition}
\textbf{\textit{Relative Importance}}. \textit{The relative importance between objective $i$ and $j$ is quantified by:}
\begin{equation}
\begin{split}
\rho_{ij} \triangleq \frac{w^i}{w^j}
\end{split}
\label{relative}
\end{equation}
\textit{\myred{which determines the locally optimal Pareto frontier location} in multi-objective optimization.}
\end{definition}
Therefore, our goal is to determine a better Pareto-optimal relative positive weights $\rho$ through weights normalization, which preserves the optimization dynamics while ensuring fair evaluation metric comparison across objectives with the same learning rate.

\subsubsection{Reinforcement Optimizing Policy}
Reinforcement learning (RL) agents are fundamentally designed to maximize cumulative rewards by interacting with environments through sequential actions. Inspired by this paradigm, we propose an Iterative Pareto Policy Optimization (IPPO) mechanism to automatically discover improved Pareto frontiers through RL principles. Notably, our ensemble ranking framework naturally aligns with RL concepts, as illustrated by the following mappings in Figure~\ref{Pantheon}(b):
\begin{itemize}[leftmargin=*,align=left]
\item \textbf{State}: The \mypurple{weights} $w^{\texttt{xtr}}>0$ indicates the locally Pareto frontier.
\item \textbf{Agent}: The trainable \mypurple{model} to chase the dynamic user interests.
\item \textbf{Environment}: Real-time user-item interaction logs for \mypurple{streaming training} and \mypurple{evaluation}.
\item \textbf{Reward}: \mypurple{Binary 0/1 signal} triggered when the reference agent outperforms the base model across all \texttt{GAUC} metrics.
\item \textbf{Action}: \mypurple{Replacing} base model, or \mypurple{adjusting} reference model's weights, e.g,. $\Delta=\frac{0.1}{N}$, where $N$ is the objectives number.
\end{itemize}

The core objective is to learn an optimal \mypurple{policy} $\pi(\texttt{Action}|\texttt{State})$ that maps states to action distributions, thereby maximizing the expected cumulative reward.
Following the self-play idea, our IPPO framework iteratively maintains two policy models: a fixed base model serving as a performance benchmark, and a reference model that explores parameter adjustments to surpass the base model.
In practice, we devise a rule-based policy governing actions on the reference model:
\begin{itemize}[leftmargin=*,align=left]
\item \textbf{If all evaluation metrics dominate}: Update the better Pareto frontier by \mypurple{replacing} the base model with the reference model.
\item \textbf{Otherwise}: \mypurple{Adjusting} the corresponding objective weight with maximum \texttt{GAUC} gap $(w^{\texttt{xtr}} = w^{\texttt{xtr}} + \Delta)$ for subsequent iterations, where $\Delta$ is small value, e.g,. $\Delta=\frac{0.1}{N}$.
\end{itemize}
In this way, we can encourage our model explores a series of `good' weights to reach a better Pareto frontier automatically.

\begin{table*}[t!]
\centering
\caption{Offline Results in term GAUC in Live-Streaming Recommendation.}
\setlength{\tabcolsep}{13pt}{
\begin{tabular}{lccccccc}
\Xhline{0.8pt} 
\textbf{Method} & \textbf{wtr} & \textbf{ltr} & \textbf{lvtr} & \textbf{ctr} & \textbf{evtr} & \textbf{inlvtr} & \textbf{inevtr}\\ 
\cline{2-8} 
\Xhline{0.4pt} 
Ranking Model & 66.49\% & 74.73\% & 69.56\% & 64.18\% & 66.07\% & 60.36\% & 60.76\%\\
\Xhline{0.4pt} 
Formulation & 57.61\% & 68.32\% & 65.81\% & 61.21\% & 60.75\% & 57.40\% & 56.97\%\\
Pantheon & 59.94\% & 71.62\% & 67.46\% & 63.76\% & 62.99\% & 58.69\% & 58.17\%\\
Improvement$\uparrow$  & \textbf{+2.33\%} & \textbf{+3.30\%} & \textbf{+1.55\%} & \textbf{+2.55\%}& \textbf{+2.24\%} & \textbf{+1.19\%} & \textbf{+1.20\%}\\
\Xhline{0.8pt} 
\end{tabular}}
\label{exp:Rank_offline_results}
\end{table*}

\section{Score Distribution Discussion}
In this section, we discuss and provide our first-hand empirical insights: how the Pantheon output distribution effect the online A/B test.
In practice, different weight group of our Pantheon will produce different fusion score distributions (in Eq.(~\ref{ensembleencoder})), thus they have different sorting abilities.
To analyze the sorting ability, we have conducted a lot of experiments and found that they have the following characteristics based on its output distributions: 
\begin{itemize}[leftmargin=*,align=left]
\item Mean value determines exposure number: larger mean value will increase the amount of live-streaming distribution to users. 
\item Variance value determines exposure position: larger variance value make users watch live-streaming earlier than short-video.
\end{itemize}
According to our observation, we could add a Mean-Variance calibration between two different models for a fair online A/B test comparison.
In our system, we align the models of the experimental Pantheon variants group to the output distribution of the baseline Pantheon. We found that this technique can ensure that the exposure numbers and positions of different models are aligned.

\begin{table*}[t!]
\centering
\caption{Online Pantheon A/B Testing Performance (\%) at Kuaihou.}
\setlength{\tabcolsep}{8pt}{
\begin{tabular}{ccccccc}
\Xhline{0.8pt} 
\textbf{Scenarios}& \textbf{Exposure}  & \textbf{Clicked User} & \textbf{Watch Time} & \textbf{In-Room Watch Time} & \textbf{Gift Count}      & \textbf{Follow}    \\ 
\Xhline{0.4pt} 
Scenario\#1  &+0.133\% &+1.010\% &+0.722\% &+0.780\% &+0.068\%  &+2.246\%     \\ \hline
Scenario\#2  &+1.927\% &+1.671\% &+1.766\% &+1.637\% &+0.202\% &+2.638\%     \\ \hline
Scenario\#3  &+0.244\% &+1.039\% &+0.257\% &+0.299\% &+1.437\%  &+1.297\%      \\ \hline
Scenario\#4  &-0.109\% &+0.518\% &+1.292\% &+0.881\% &+1.335\% &+0.208\%     \\ \hline
\Xhline{0.8pt} 
\end{tabular}
}
\label{tab:Online_results}
\end{table*}

\begin{table*}[t!]
\centering
\caption{Offline Ablation Studies in term of GAUC at Live-streaming recommendation.}
\setlength{\tabcolsep}{12pt}{
\begin{tabular}{lccccccc}
\Xhline{0.8pt} 
\textbf{Method} & \textbf{wtr} & \textbf{ltr} & \textbf{lvtr} & \textbf{ctr} & \textbf{evtr} & \textbf{inlvtr} & \textbf{inevtr}\\ 
\cline{2-8} 
\Xhline{0.4pt} 
Formulation & 57.61\% & 68.32\% & 65.81\% & 61.21\% & 60.75\% & 57.40\% & 56.97\%\\
\Xhline{0.4pt}
Pxtr\&MLP &59.20\% &70.93\% &66.91\% &63.23\% &62.13\% &57.92\% &57.81\% \\
Hidden-State\&MLP & 59.94\% & 71.62\% & 67.46\% & 63.76\% & 62.99\% & 58.69\% & 58.17\%\\
Hidden-State\&Transformer &60.56\% &72.15\% &67.57\% &63.89\% &63.06\% &58.79\% &58.31\%\\
\Xhline{0.8pt} 
\end{tabular}}
\label{exp:abulation}
\end{table*}

\section{Experiments}
In this section, we answer the following research questions:
\begin{itemize}[leftmargin=*,align=left]
    \item \textbf{RQ1}: How does Pantheon bring gains in offline evaluation?
    \item \textbf{RQ2}: How does Pantheon contributes online improvements?
    \item \textbf{RQ3}: What is the impact of model architecture and model input on offline performance?
    \item \textbf{RQ4}: How does Pantheon changes our services ecology?
    \item \textbf{RQ5}: How does the final fusion score more balanced in its dependence on multiple objectives?
\end{itemize}

\subsection{Data-streaming and Metrices}
\label{gauc_exp}
We conduct extensive experiments at the live-streaming recommendation services at Kuaishou, which is one of the largest recommendation scenario including over 400 million users and several billion exposure logs in our data-streaming.
For evaluation, we select the the wide-used \texttt{GAUC} as our evaluation metric, since it can comprehensively measure the model's ability among users:
\begin{equation}
\begin{split}
\texttt{GAUC} = \sum_{u} w_u \texttt{AUC}_u \quad \texttt{where}\ \ w_u = \frac{\texttt{exposure}_u}{\texttt{all exposure}},
\end{split}
\label{gauc}
\end{equation}
where the $w_u$ denotes the user $u$'s exposure ratio, the $\texttt{AUC}_u$ is the AUC-ROC results of user $u$.

\subsection{Offline Comparison (RQ1)}
The main experiment results are shown in Table~\ref{exp:Rank_offline_results}, which describes the most importance objectives evaluation results in our system (i.e., \textbf{wtr}/follow, \textbf{ltr}/like, \textbf{lvtr}/long-view, \textbf{ctr}/click, \textbf{evtr}/effective-view, \textbf{inlvtr}/in-room long-view and \textbf{inevtr}/in-room effective-view).
Here we first report the Ranking model metrics, since it estimates each objective separately, thus it represents the \textbf{ceiling performance} of our system.
Based on the Ranking model output \texttt{Pxtrs}, we next show our online formulation results, which have been served as our ensemble sort stage solution for past several years.
From it, we could found that the offline evaluation results are reduced under various objectives, e.g., $66.4\% \Rightarrow 57.6\%$ in wtr.
It is reasonable to have a such performance degradation since the trade-off between performance and interpretability in multi-objective optimization.
For our Pantheon, it obviously shows statistical improvements over the formulation-based ensemble sort in all objectives with a large number of improvements, \textbf{average +1.62\% in different objectives}.
It is the largest modification in past years at Kuaishou live-streaming, which demonstrates our Pantheon could automatically converge to a better state to balance different objectives' performance under the proposed IPPO technique.

\subsection{Online A/B Tests (RQ2)}
In this section, we deploy our Pantheon to four different Scenarios to replace the previous formulation-based ensemble sort to response real recommendation requests.
In our service, the most important online metrics are the clicked user, watching-time and gift count,  which reflect the watching live-streaming user group, total amount of time spend on live-streaming and the value of digital gifts.
Besides, we also report the exposure metric, which measures the load of live-streaming users watched, to help us make a fair comparison with the base formulation-based ensemble sort approach.
Specifically, due to the large scale of our business, the about 0.1\% improvement in clicked User is statistically significant enough to our system.
According to Table~\ref{tab:Online_results}, we could find that our Pantheon achieves a very
significant improvement of +1.010\%, +1.671\%, +1.039\% and +0.518\% in term of the clicked user in four different scenarios, respectively.
Meanwhile, our Pantheon also bring significant gain at the watch time and interaction metrics (the biggest gains in past year), which reveals our method could converge to a better Pareto frontier for our system.

\subsection{Ablation Study (RQ3)}
In this section, we construct several model variants to validate our Pantheon framework effectiveness.
As shown in Figure~\ref{Pantheon}(a), our Pantheon has two additional optimizing directions to enhance ensemble sort ability, from the model input (in Eq.(\ref{input})) and ensemble encoder (in Eq.(\ref{ensembleencoder})).
For a fair comparison, we conducted three aspects modifications:
\begin{itemize}[leftmargin=*,align=left]
\item \textbf{Pxtr v.s. Hidden-State}: Does the high-dimensional hidden-state \textbf{t} preserves more information than the highly-compressed \texttt{Pxtr}, thereby improving ensemble sort performance?
\item \textbf{MLP v.s. Transformer}: Does the advanced model architecture could further enhance ensemble sort performance?
\end{itemize}
According to Table~\ref{exp:abulation}, we can draw the following conclusions:
(1) Compared with Formulation, the Pxtr\&MLP shows superior performance across these objectives.
It demonstrates the effectiveness of our IPPO mechanism to ensure our framework can be convergence to a promising Pareto frontier adaptively.
(2) Compared with Pxtr\&MLP, the Hidden-State\&MLP consistently yields the significant accuracy improvement on all objectives, which reveals that joint training our Pantheon with Ranking model's multi-task module hidden-state is a more powerful way to enhance ensemble sort ability
(3) Instead of utilizing the MLP as the \texttt{Ensemble\_Encoder}$(\cdot)$, we further upgrade it to the Transformer architecture to obtain the final score. Compared with Hidden-State\&MLP, the Hidden-State\&Transformer version shows significant gains over MLP-based Pantheon, which demonstrates a new direction to explore more complex architecture to build a more powerful ensemble sort.

\begin{table}[t!]
    \centering
    \caption{Exposure Ecology Analysis across User Group.}
    \setlength{\tabcolsep}{8pt}{
    \begin{tabular}{lccc}
    \Xhline{0.8pt}
    User Group & Activate & Exposure & In-Room Eff-View\\ 
        \Xhline{0.4pt}
        \multirow{3}{*}{In-Room} & High &-2.2\% &+3.5\%\\
         & Mid& -0.9\% &+5.7\%\\
         & Low& +3.6\% &+9.8\% \\
         \Xhline{0.4pt}
         \multirow{3}{*}{Out-Room} & High &+0.7\% &+4.8\% \\
         & Mid& +6.4\% &+12.9\%\\
         & Low& +4.3\% &+6.9\%\\
    \Xhline{0.8pt}
    \end{tabular}}
\label{exp:exposureecology}
\end{table}

\begin{table}[t!]
    \centering
    \caption{Behaviour Pattern Ecology Analysis.}
    \setlength{\tabcolsep}{10pt}{
    \begin{tabular}{lcc}
    \Xhline{0.8pt}
    Classification & Behaviours & Improvement \\ 
        \Xhline{0.4pt}
        \multirow{4}{*}{Multiple-Good} & Long\&Inter\&Gift &+0.83\% \\
         & Long\&Inter& +5.32\%\\
         & Long\&Gift& +0.53\%\\
         & Inter\&Gift& +3.65\%\\
         \Xhline{0.4pt}
         \multirow{3}{*}{Single-Good}& Long-View& +10.84\%\\
         & Interaction& +9.13\%\\
         & Gift& +1.87\%\\
    \Xhline{0.8pt}
    \end{tabular}}
\label{exp:behaviourecology}
\end{table}

\begin{table}[t!]
    \centering
    \caption{Kendall's $\tau$ Coefficient Analysis.}
    \setlength{\tabcolsep}{4.5pt}{
    \begin{tabular}{lcccc}
    \Xhline{0.8pt}
    \multirow{2}{*}{\makecell[c]{\textbf{Pxtr of}\\\textbf{Ranking Model}}} & \multicolumn{2}{c}{Scenario\#1} & \multicolumn{2}{c}{Scenario\#2} \\  & Fomula & Pantheon & Fomula & Pantheon \\ 
        \Xhline{0.4pt}
        \textbf{inlvtr} &0.2657 &0.2158 &0.2278 &0.2106 \\
        \textbf{inetr} &0.2230 &0.2104 &0.2001 &0.2009 \\
        \textbf{lvtr} &0.2099 &0.1412 &0.1898 &0.1339 \\
        \textbf{evtr} &0.1645 &0.1297 &0.1514 &0.1197 \\
        \textbf{ltr} &0.1194 &0.1011 &0.1186 &0.1032 \\
        \textbf{wtr} &0.1054 &0.1023 &0.0997 &0.1044 \\
    \Xhline{0.8pt}
    \end{tabular}}
\label{exp:kendall}
\end{table}

\subsection{Ecology Changes (RQ4)}
In this section, we aim to answer a question: what kind of changes does our Pantheon ensemble sort bring to our service ecology?
As shown in Figure~\ref{kuaishou}, live-streaming has two styles (e.g. Out-Room/In-Room form), and different users have different habits when they surf on our platform.
For the diverse users, we first divide them into two user group categories: those who have a preference for watching In-Room and those who usually watch Outside-Room.
Regarding the two different user group, we further divide them into three groups according to whether they are active on our platform.
Here we show empirical performance in Table~\ref{exp:exposureecology}, which describes the exposure changes rate compared with the formulation-based ensemble sort method, and we have the observations:
\begin{itemize}[leftmargin=*,align=left]
\item The exposure traffic has shifted from "In-Room users" to "Out-Room users", while it has increased the effective usage scale. This phenomenon shows that our Pantheon optimizes traffic exposure from the top user group, frees up capacity for potential new users, and increases opportunities to distribute to more users.
\item For the In-Room user group, our Pantheon utilizes a smaller traffic pool but drives a higher experience, e.g., In-Room Eff-View rate +4.8\%/+12.9\%/+6.9\%. This phenomenon shows that our Pantheon recommends more right live-streamings to our users, building a
better environment for our system.
\end{itemize}

Moreover, to investigate the impact of our Pantheon effects more comprehensively, we conduct another comparison from the user behaviour perspective.
From the perspective of live-streaming business, we can classify the user's behavior into three aspects (i.e., watching time, interaction, gift) to identify good user watching.
\begin{itemize}[leftmargin=*,align=left]
\item Single-Good: only one of the following signals occurred during user watching a live-streaming: long-view (> 60s), interaction (like/comment/forward/follow), or send gift.
\item Multiple-Good: occurs two or more signals.
\end{itemize}
There we show the ecology change results in Table~\ref{exp:behaviourecology}, our Pantheon  consistently yields significant improvements across the single-good and multiple-good behaviours.
We suppose the reason lies in that our IPPO follows the Pareto optimizing strategy, which ensure that optimizing one metric without reducing other metrics.

\subsection{Objective Dependence Analysis (RQ5)}
In this section, we explore the fusion score's dependency among the different objectives.
Here we conducted the Kendall's $\tau$ analysis between the \textbf{Ranking model} \texttt{Pxtr} with the fusion scores of formula and our pantheon. 
Specifically, the Kendall's $\tau$ coefficient value is in the ranges from $[-1, +1]$, higher score means more related.
According to the Table~\ref{exp:kendall}, we could find that our Pantheon also maintains the \textbf{exact same order of importance among the multiple objective} with the origin formula, while has the minor variations in the absolute magnitude of the coefficient weights.
Such phenomenon demonstrates that our Pantheon  is capable of effectively capturing a more balanced objectives usage.
 We attribute the reason to our model utilizing the high-dimensional representations, instead of the highly-compressed numerical scores \texttt{Pxtr}, which could enhance our ensemble sort robustness to reduce the numerical fluctuations caused by top important objectives.

\section{Conclusion}
In this paper, we explore a more flexible ensemble sort approach, Pantheon, which fully replaced the wide-used formula-based ensemble sort and successfully deployed at Kuaishou, serving 400 Million users per day.
Specifically, compared with traditional formula-based ensemble sort, our Pantheon has several key insights:
(1) our Pantheon is a `plugin' component that jointly training with Ranking model. Meanwhile, we do not utilize the highly-compressed numerical scores Pxtr as input, but employ the high-dimensional task representations as Pantheon inputs, which maintains more fine-grained information.
(2) Our Pantheon follows a reinforcement optimizing policy IPPO to search the most appropriate objective weights automatically, which could 
guarantee our model to find better Pareto frontier without hurt some objectives to improve other objectives.
In the future, we will utilize our Pantheon as reward model, to explore industrial-scale generative recommendation.

\balance
\bibliographystyle{ACM-Reference-Format}
\bibliography{sample-base-extend.bib}
\end{document}